\documentclass[acmsmall]{acmart}
\usepackage{graphicx}
\usepackage{hyperref}
\usepackage{amsmath}
\usepackage{subcaption}
\usepackage{pifont}

\usepackage{amssymb} 
\usepackage{color}
\usepackage{longtable}
\usepackage{multirow}
\usepackage{booktabs}
\usepackage{array}
\usepackage{algorithm,algorithmic}

\AtBeginDocument{%
 \providecommand\BibTeX{{%
 \normalfont B\kern-0.5em{\scshape i\kern-0.25em b}\kern-0.8em\TeX}}}

\begin{document}

\title{Personalized Inhibition Training with Eye-Tracking: Enhancing Student Learning and Teacher Assessment in Educational Games}

\author{Abdul Rehman}
\email{arj@hvl.no}
\affiliation{
 \institution{Department of Computer Science, Electrical Engineering and Mathematical Sciences, Western Norway University of Applied Sciences}
 \city{Bergen}
 \country{Norway}
}

\author{Ilona Heldal}
\email{ilona.heldal@hvl.no}
\affiliation{
 \institution{Department of Computer Science, Electrical Engineering and Mathematical Sciences, Western Norway University of Applied Sciences}
 \city{Bergen}
 \country{Norway}
}

\author{Diana Stilwell}
\email{dianateixeira@edu.ulisboa.pt}
\affiliation{
 \institution{CICPSI, Faculdade de Psicologia, Universidade de Lisboa}
 \streetaddress{Alameda da Universidade 1649-013}
 \city{Lisboa}
 \country{Portugal}
}

\author{Paula Costa Ferreira}
\email{paula.ferreira@edu.ulisboa.pt}
\affiliation{
 \institution{CICPSI, Faculdade de Psicologia, Universidade de Lisboa}
 \streetaddress{Alameda da Universidade 1649-013}
 \city{Lisboa}
 \country{Portugal}
}

\author{Jerry Chun-Wei Lin}
\email{jerry.chun-wei.lin@hvl.no}
\affiliation{
 \institution{Department of Computer Science, Electrical Engineering and Mathematical Sciences, Western Norway University of Applied Sciences}
 \city{Bergen}
 \country{Norway}
}

\renewcommand{\shortauthors}{Rehman et al.}

\begin{abstract}

Eye tracking (ET) can help to understand visual attention and cognitive processes in interactive environments. This study presents a comprehensive eye-tracking analysis framework of the Inhibitory Control Game, named the ReStroop game, which is an educational intervention aimed at improving inhibitory control skills in children through a recycling-themed sorting task, for educational assessment that processes raw gaze data through unified algorithms for fixation detection, performance evaluation, and personalized intervention planning. The system employs dual-threshold eye movement detection (I-VT and advanced clustering), comprehensive Area of Interest (AOI) analysis, and evidence-based risk assessment to transform gaze patterns into actionable educational insights. We evaluated this framework across three difficulty levels and revealed critical attention deficits, including low task relevance, elevated attention scatter, and compromised processing efficiency. The multi-dimensional risk assessment identified high to moderate risk levels, triggering personalized interventions including focus training, attention regulation support, and environmental modifications. The system successfully distinguishes between adaptive learning and cognitive overload, providing early warning indicators for educational intervention. Results demonstrate the system's effectiveness in objective attention assessment, early risk identification, and the generation of evidence-based recommendations for students, teachers, and specialists, supporting data-driven educational decision-making and personalized learning approaches.
\end{abstract}




\maketitle

\section{Introduction}
Eye tracking (ET) is a widely used technology that analyzes how individuals visually interact with digital environments through gaze measures \cite{rehman2024towards,novak2024eye,ali2023towards}. By measuring where and for how long someone focuses on different areas of a screen, valuable insights can be gained into attention, decision-making processes, and cognitive load \cite{daehlen2024towards}. Recently, ET technology has been significantly applied in serious game-based scenarios, where users dynamically engage with evolving visual representations \cite{costescu2023mushroom}. Unlike static images, games introduce variation in visual attention during gameplay \cite{costescu2023mushroom}, which can be recorded and analyzed through gaze data collected by ET. Analyzing changes in gaze over time, along with the progression of games and user interactions, is referred to as temporal analysis \cite{lamsa2022focus}. This approach enables us to not only identify the areas where people direct their gaze but also to approximate attention transitions throughout the gaming experience \cite{keshava2024just}.

The Inhibition game is a 3D educational environment designed to assess and enhance cognitive inhibitory control through interactive trash sorting tasks across multiple difficulty levels \cite{diamond2013executive, zelazo2003development}. Players navigate three progressive stages, where they must categorize various objects such as glass, plastic, paper, and organic waste into their corresponding recycling bins while resisting visual distractions and ignoring irrelevant stimuli \cite{rueda2005training, klingberg2010training}. The game employs a dual-task paradigm, requiring students to maintain focus on correct object-bin associations while inhibiting responses to misleading visual cues, incorrect bin placements, and time pressure elements \cite{miyake2000unity, friedman2006unity}. Each stage increases cognitive load by introducing more complex sorting rules, additional distractor objects, and reduced response time windows, effectively measuring the development of executive function skills \cite{best2011relations, blair2011measuring}. The 3D environment captures real-time eye-tracking data, including gaze coordinates, object interaction patterns, and definitions of areas of interest (AOI), providing rich behavioral data for analyzing attention allocation and inhibitory control strategies \cite{holmqvist2011eye, rayner2009eye}.

\textbf{Motivation: } The motivation for developing this eye-tracking insights system arises from the pressing need for objective, real-time assessment tools that can assist both students and teachers in understanding cognitive development patterns \cite{anderson2002development}. Visual attention and inhibitory control are fundamental cognitive processes that have a significant impact on academic performance and learning outcomes in educational settings. Traditional assessment methods for measuring these cognitive abilities often rely on subjective observations or limited standardized tests, which fail to capture the dynamic and real-time nature of attention regulation during task performance. To address these challenges, it is essential to create objective, data-driven assessment tools for the early identification of attention-related learning difficulties and to enable targeted educational interventions. Eye-tracking technology offers unprecedented opportunities to quantify visual attention patterns, cognitive processing strategies, and inhibitory control mechanisms through precise measurement of gaze behavior during task execution. Inhibition games (ReStroop), specifically designed to assess cognitive control and attention regulation abilities, have emerged as valuable tools for both educational and clinical assessment. Research by Diamond \cite{diamond2013executive} shows that inhibitory control tasks effectively measure the development of executive functions within academic contexts.
In contrast, studies by Rueda et al. \cite{rueda2005training} demonstrate that attention training through inhibition-based games can significantly improve cognitive control abilities in children. Recent work by Posner and Rothbart \cite{posner2007attention} underscores the critical role of attention networks in academic achievement, emphasizing the necessity for objective measurement tools that can capture individual differences in attention regulation. Furthermore, research by Klingberg \cite{klingberg2010training} indicates that targeted attention training interventions based on objective assessments can lead to sustained improvements in academic performance. However, existing assessment approaches lack the comprehensive and real-time analysis capabilities required to generate personalized intervention strategies based on detailed cognitive processing patterns, highlighting a critical gap that this eye-tracking analysis system aims to fill.

\textbf{Contributions:} This framework provides a comprehensive framework for assessing attention in educational games using eye-tracking technology. A significant technical advancement involves solving the 3D-to-2D coordinate transformation problem, which enables precise detection of Areas of Interest (AOI) in complex game environments. Building on this foundation, the system delivers real-time analysis of attention metrics, including AOI focus, content engagement, and resistance to distraction, all through a fully automated pipeline. This is the first system to extract insights from dual perspectives, generating personalized feedback for both students and teachers based on the same eye-tracking data. An integrated classification algorithm categorizes learners into three profiles: Advanced, Developing, or Struggling, based on their attention patterns, thus laying the groundwork for adaptive learning. The system also supports longitudinal tracking of attention development across different levels of game difficulty, providing a dynamic view of learning progression. Teachers benefit from early intervention tools that help identify students at risk before they encounter academic challenges. Meanwhile, students gain self-awareness through age-appropriate visualizations of their attention behaviors and learning strategies. Finally, a comprehensive validation framework ensures both the technical accuracy and educational effectiveness of the system, making it a robust solution for cognitive assessment and support in classroom settings.

The remainder of this paper is organized as follows: Section~\ref{sec_relatedwork} discusses existing work related to inhibition control. Section~\ref{sec_framework} introduces the proposed inhibition framework and details the methodology, including parameter selection (Section~\ref{sec_params}) and dataset selection along with the pre-processing steps (Section~\ref{sec_dataset}). Section~\ref{sec_Experimental} outlines the experimental setup and presents the results and their analysis. Finally, Section~\ref{sec_conclusion} wraps up the paper and highlights future directions for research.

\section{Related Work}\label{sec_relatedwork}

Frutos-Pascual et al. \cite{frutos2015assessing} utilized eye-tracking (ET) technology to assess children's behavior in attention-enhancement therapies. By analyzing eye movement patterns during interactions with puzzle games, the researchers discovered that participants who performed better displayed quantifiably different eye movement patterns compared to those with poorer results. Piazzalunga et al. \cite{piazzalunga2023investigating} employed serious games-based ET data to identify children at risk of dysgraphia. Their analysis of scan paths and fixation patterns during gameplay revealed that children with lower performance exhibited chaotic scan paths, whereas those who performed better demonstrated more organized scan paths. The integration of ET metrics with game performance data provided a detailed understanding of visual perception impairments. Velichkovsky et al. \cite{velichkovsky2019visual} examined the distribution of fixation durations among professional, amateur, and novice eSports players. They found that highly skilled gamers exhibited more variability in fixation durations and displayed bimodal distribution patterns, indicating the presence of both ambient and focal fixation types. Argasinski et al. \cite{argasinski2017patterns} proposed a framework for designing and evaluating serious games that incorporate ET and biosensor data. They suggested that including temporal data enhances the assessment of user interactions and affective responses during gameplay. Hajari et al. \cite{hajari2018spatio} explored team cognition by analyzing spatio-temporal ET data during laparoscopic simulation operations. Using Cross Recurrence Analysis (CRA) and overlap analysis, they identified features that differentiate high-performing teams from low-performing ones based on temporal gaze data patterns, thereby enriching the understanding of collaborative performance.

Diamond et al. \cite{diamond2013executive} outlined a framework for understanding executive functions, mental skills such as working memory, cognitive flexibility, and inhibitory control critical for learning and success throughout life. Her research reveals that these functions are more indicative of school readiness than IQ and can be enhanced through targeted interventions. This framework helps explain how children develop self-regulation skills and why some struggle with attention and behavioral control in educational settings. Zelazo et al. \cite{zelazo2003development} analyzed the emergence of executive functions in early childhood, emphasizing the importance of the preschool years. Their findings highlight rapid development of these skills between ages 3 and 7, particularly inhibitory control, which is essential for academic success. They also identify key developmental periods for effective interventions. Rueda et al. \cite{rueda2005training} demonstrated that executive attention can improve through targeted computerized training, leading to measurable gains in children's conflict resolution and sustained attention. Their research supports the idea that executive functions are malleable and can be developed through practice. These findings suggest that technology-based training programs could be practical tools for helping children with focus and self-regulation challenges.

Klingberg et al. \cite{klingberg2010training} demonstrated that working memory can be improved through adaptive training, leading to structural brain changes and lasting effects on cognitive abilities and academic performance. Their findings suggest that such training could benefit students with attention and learning difficulties. Miyake et al. \cite{miyake2000unity} established a framework for understanding executive functions, highlighting that they include distinct yet related components: inhibition, shifting, and updating. This research indicates that interventions should target specific aspects of executive function for better results in learning. Friedman et al. \cite{friedman2006unity} refined the understanding of executive functions by examining the relationship between inhibition and interference control, revealing that these components have both shared and unique aspects. Their work underscores the need for comprehensive evaluation methods to assess individual differences in self-regulation abilities effectively. Best et al. \cite{best2011relations} conducted a significant longitudinal study showing that executive function abilities are linked to academic achievement from early childhood through adolescence. Their research demonstrates that skills assessed as early as age 5 can predict future academic outcomes, particularly in mathematics, where inhibitory control plays a crucial role. The study emphasizes the need to support executive function development, as it contributes to ongoing academic success. Additionally, interventions targeting these skills could greatly benefit students' educational paths.

% Best et al., \cite{best2011relations} conducted a large-scale longitudinal study demonstrating strong relationships between executive function abilities and academic achievement across development from early childhood through adolescence. Their research shows that executive functions measured as early as age 5 predict academic outcomes years later, with inhibitory control being significant for mathematics achievement. The study provides compelling evidence that executive function skills are not just related to school readiness but continue to influence academic success throughout the school years. The findings highlight the importance of supporting executive function development as a means of promoting long-term educational success. This research emphasizes that interventions targeting executive functions could have broad benefits for students' academic trajectories.

\section{Inhibition Framework (Proposed Tool)}
\label{sec_framework}
\begin{figure}[!ht]
 \centering
 \includegraphics[width=\linewidth]{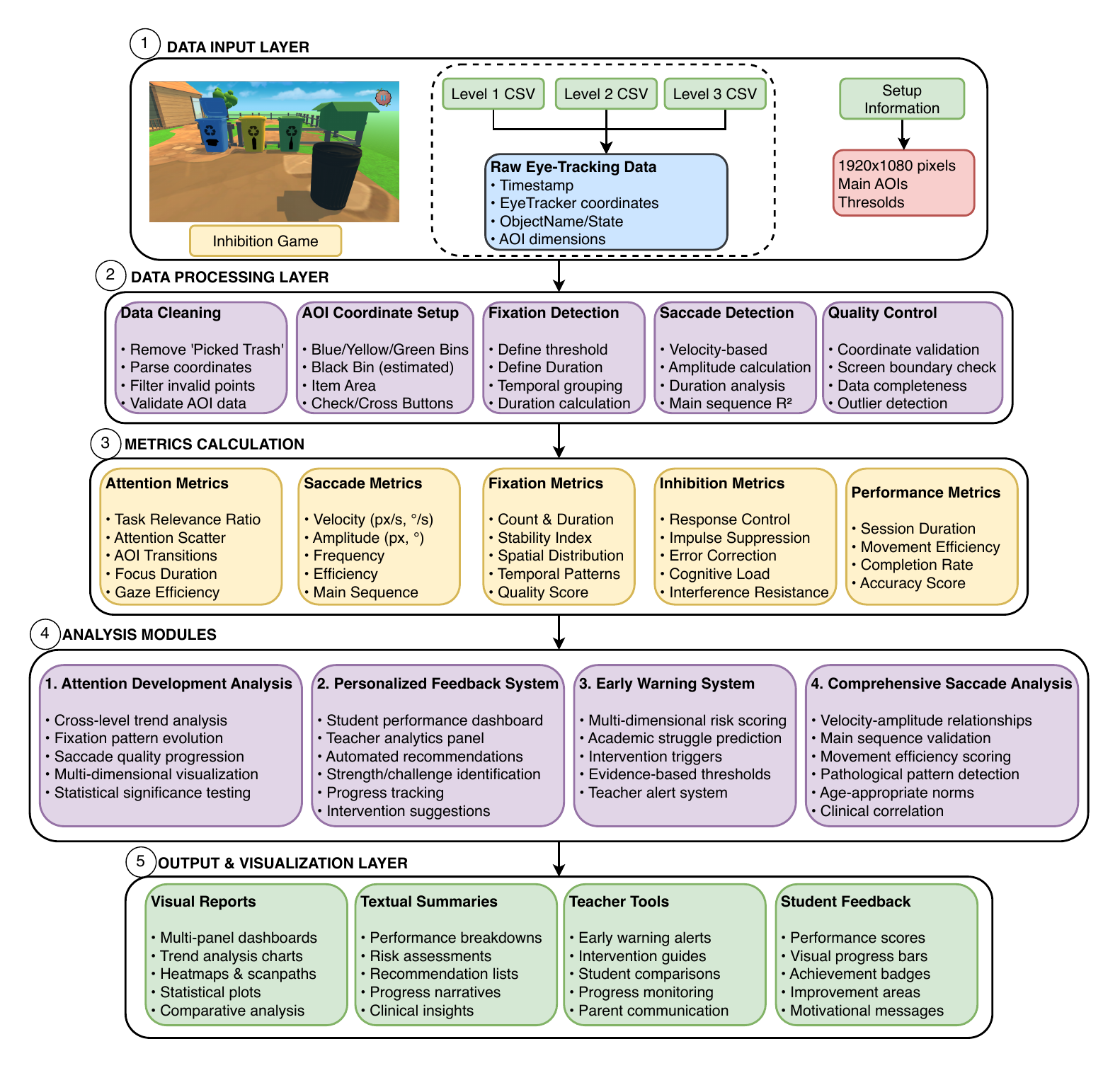}
 \caption{Proposed tool for Time-Series Analysis of Eye-Tracking Data}
 \label{figpropsod}
\end{figure}

Figure \ref{figpropsod} illustrates the proposed tool, which encompasses various stages: data collection, processing, quality validation, parameter selection, gaze classification, and performance metrics. It addresses two scenarios: a first-level overall analysis for detailed insights and a multilevel analysis to compare performance across three levels. This approach effectively transforms raw gaze coordinates into actionable educational insights through objective measurement of attention, early identification of risks, and personalized learning support. The system facilitates data-driven educational decision-making by offering comprehensive cognitive assessments and evidence-based intervention recommendations.

\subsection{Dataset Selection and Pre-Processing} \label{sec_dataset}
This study utilizes the \textbf{ReStroop game}, an inhibition training tool, as part of the larger EU-funded EMPOWER project \cite{projectempowerProjectEMPOWER}. The inclusion of eye-tracking data collection was prioritized in all games during this project phase. To streamline session length and enhance the reliability of our dataset, gaze data collection was limited; it was only gathered during the sustained attention game. Our dataset is based on data collected through the EMPOWER project (grant agreement No. 101060918), which features nine games designed to teach regulation strategies for various cognitive functions. The ReStroop game itself is an educational intervention aimed at improving inhibitory control skills in children through a recycling-themed sorting task. Drawing on the classic Stroop paradigm, players must categorize waste items into the correct recycling bins while managing conflicting visual information regarding the colors of the objects and bins. This cognitive challenge specifically targets the ability to suppress automatic responses that may conflict with task demands, a fundamental aspect of executive function.

The game consists of twelve progressively difficult levels, with advanced stages introducing time constraints. Players receive immediate feedback on their performance through a comprehensive scoring system. They are rewarded with bronze, silver, or gold badges based on their accuracy percentages (25-49\%, 50-74\%, and 75-100\%, respectively). Additionally, players can earn stars that can be redeemed in an integrated reward store. This gamified approach to cognitive training combines environmental education with the development of executive function skills. It provides valuable insights into recycling practices while promoting systematic practice of inhibitory control processes essential for academic and behavioral self-regulation. There is also a Teacher Platform that allows educators to review student data, monitor progress, and plan interventions. Teachers first select the language and then connect to the server to view active interventions. They can create new intervention plans tailored for individual students, and as students progress through the games, their progress is displayed on the screen.  The platform features an interface where teachers can see which levels students have completed, what remains, and any errors students have made. Teachers can also provide intervention plans for a specified number of weeks or sessions. Once all steps are completed, feedback is generated for each student. The gameplay mechanics of the Re Stroop task involve two distinct experimental conditions that manipulate cognitive conflict to assess inhibitory control.

In the congruent condition, participants encounter scenarios where the waste item's material properties match the color of the corresponding bin. For example, a green plastic bottle appears when the green recycling bin is open. This condition requires minimal inhibitory control, as the visual cues support the correct response, allowing participants to select the checkmark to confirm the match. In contrast, the incongruent condition presents conflicting information, where the waste item’s material does not match the available bin color. For instance, a yellow cardboard box appears when only the blue trash bin is accessible. This mismatch creates cognitive interference, demanding active inhibition of the automatic response tendency. In this case, participants must select the X symbol to reject the inappropriate pairing. This experimental design parallels the classic Stroop effect, where participants must override automatic responses based on superficial visual similarities in favor of a deeper categorical analysis. The systematic alternation between congruent and incongruent trials provides a controlled environment for measuring individual differences in executive function, while also maintaining ecological validity through the real-world context of waste sorting and environmental responsibility.

\begin{figure*}[!ht]
 \centering
 \begin{subfigure}[b]{0.49\linewidth}
 \centering
 \includegraphics[width=\linewidth,height=5cm]{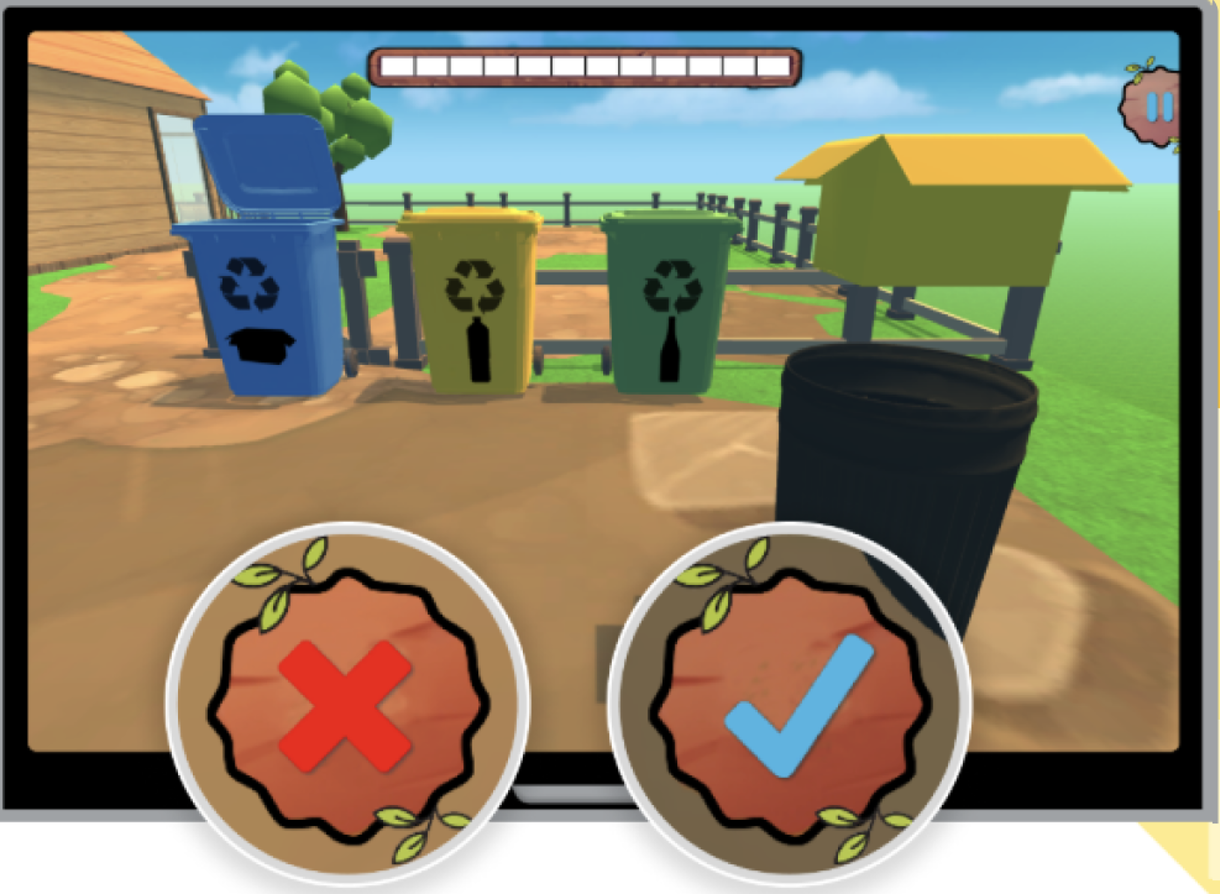}
 \caption{Game Interface (1)}
 \label{fig:GameInterface}
 \end{subfigure}
 \begin{subfigure}[b]{0.49\linewidth}
 \centering
 \includegraphics[width=\linewidth,height=5cm]{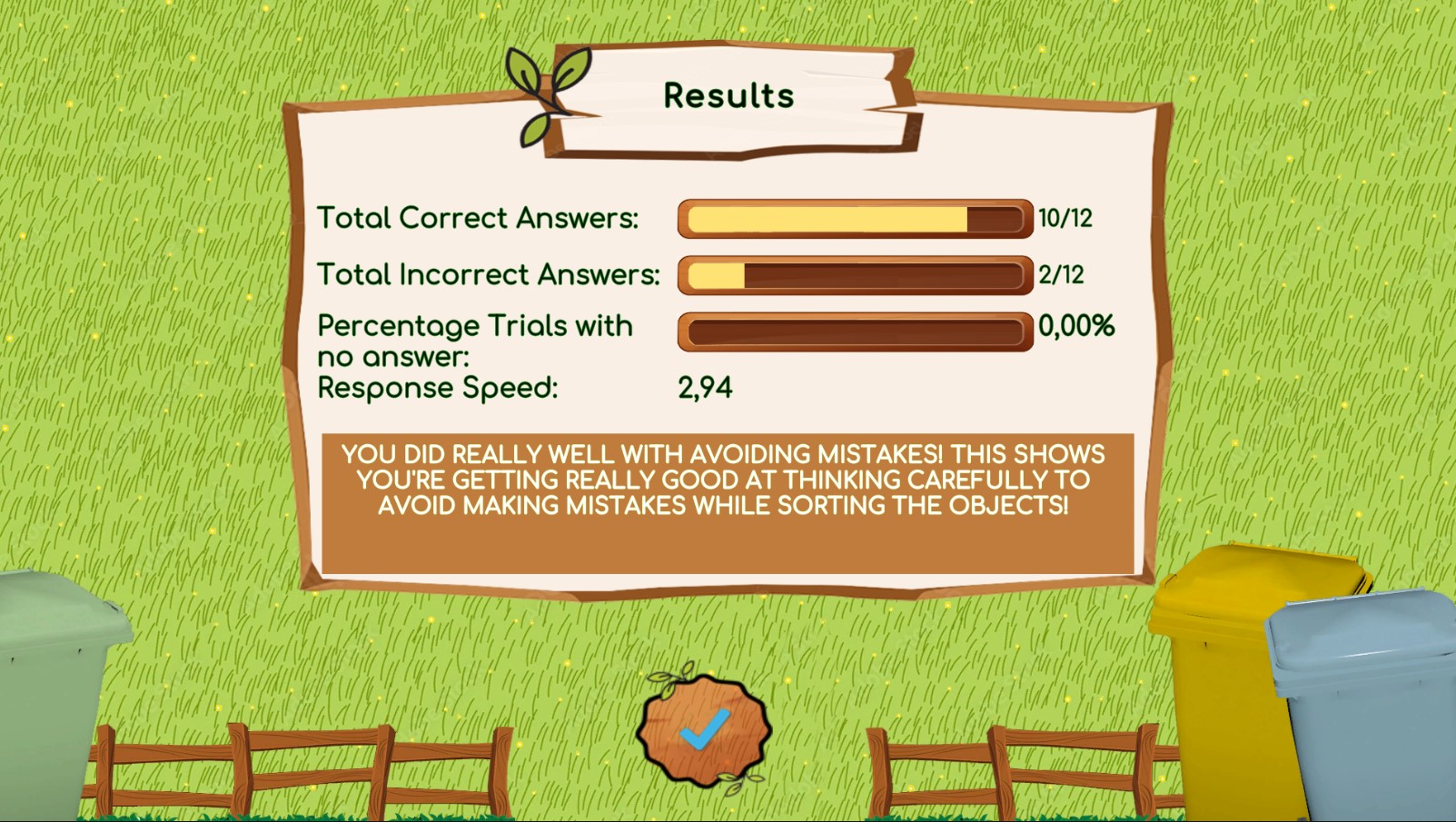}
 \caption{Game Interface (2)}
 \label{fig:GameInterface1}
 \end{subfigure}
 \caption{Game Interface of Inhibition Game}
 \label{fig:Gui}
\end{figure*}

Before analyzing attention patterns, we first clean and prepare the raw eye-tracking data to ensure accurate results. The preprocessing begins by loading CSV files from the 3D inhibition game and converting coordinate information from text strings into numerical values. This process extracts gaze positions (x, y), object locations, and attention region boundaries from the stored data. One of the most critical steps is addressing coordinate system mismatches. The 3D game employs different positioning rules compared to the 2D screen, where eye movements are measured. This involves scaling normalized coordinates (ranging from 0 to 1) to actual screen pixels, inverting Y-coordinates when necessary (since 3D games often measure from the bottom-up while screens measure from the top-down), and expanding tiny attention regions that are too small for reliable gaze detection to a minimum size of 80x80 pixels. Next, we remove invalid data points, such as (0,0) coordinates, which indicate tracking failures, and filter out gaze points that fall outside the screen boundaries. We also ensure all attention regions fit correctly within the display area. This comprehensive preprocessing transforms messy raw data into clean, reliable measurements, laying the groundwork for further attention, analysis and insights generation. Fig. \ref{fig:GameInterface} and \ref{fig:GameInterface1} depict the game interface.

\subsection{Parameter Selection}\label{sec_params}
The selection of relaxed parameters for eye movement detection is fundamentally motivated by the inherent variability in oculomotor behavior across individuals, experimental conditions, and data quality characteristics, particularly when working with NDD populations engaging with interactive digital environments such as serious games. Traditional fixed-threshold approaches suffer from significant limitations, as demonstrated by Andersson et al. \cite{andersson2017one}, who showed that slight variations in velocity and dispersion thresholds can dramatically affect detection performance, with optimal parameters varying substantially across different experimental paradigms and participants. This variability becomes even more pronounced in NDD populations, where atypical eye movement patterns, increased fidgeting, and reduced compliance with experimental instructions can lead to considerably noisier data compared to neurotypical adults. Individual differences in baseline eye movement characteristics further compound the challenge. Komogortsev et al. \cite{komogortsev2010standardization} demonstrated that individualized velocity and acceleration thresholds can significantly improve detection accuracy compared to population-based fixed thresholds. In the context of serious games for NDD children, these individual differences are amplified by the dynamic nature of the gaming environment, where participants may exhibit varying levels of engagement, attention, and motor control throughout the session.

The need for an adaptive approach to eye movement analysis has been effectively addressed by \cite{nystrom2010adaptive}, who developed an innovative velocity threshold algorithm. This algorithm is designed to be "less sensitive to variations in noise level and settings-free for the user," which is especially important when working with populations where data quality can fluctuate significantly both within and between sessions. Furthermore, varying noise levels across different eye-tracking systems and experimental setups necessitate flexible parameter adaptation. This requirement is underscored by \cite{engbert2006microsaccades}, whose algorithm automatically adjusts detection thresholds based on the identified noise level of the data while compensating for missing or corrupted samples. Such issues are particularly prevalent when tracking the eye movements of children with neurodevelopmental disorders (NDDs), who may frequently look away from the screen or engage in head movements during gameplay. The shift toward machine learning approaches, as exemplified by \cite{zemblys2018using}, directly confronts the fundamental limitation that "parameters have to be adjusted based on eye movement data quality." This transition eliminates the need for manual parameter tuning through data-driven classification, which is invaluable in serious games research due to the inherent complexity of visual stimuli and the unpredictability of participant behavior that render traditional parameter optimization impractical.
Additionally, \cite{hessels2017noise} introduced the I2MC algorithm, specifically designed to handle "data across a wide range of noise levels and periods of data loss," addressing the unique challenges faced by pediatric NDD populations prone to data loss and high noise levels. Collectively, these studies emphasize that relaxed and adaptive parameter selection is not merely a methodological preference but a fundamental necessity in response to the intrinsic complexity and variability of human eye movement data. This approach ensures robust detection performance across diverse experimental conditions and participant populations, particularly when investigating NDD children's interactions with dynamic and engaging severe game environments. Such settings present unique methodological challenges that require a careful balance between maintaining ecological validity and preserving data quality.

%\textcolor{red}{Fixation potins need to update in code, 50 px}

Algorithm~\ref{preprocessing} outlines the data preprocessing steps necessary to create a clean data foundation. This includes parsing gaze coordinate strings, filtering out invalid measurements, and establishing a comprehensive Area of Interest (AOI) framework for recycling game elements. The spatial mapping facilitates accurate attention tracking within the relevant regions of the interface.

\begin{algorithm}[!ht]
\caption{Data Pre-processing and Spatial Framework}
\label{preprocessing}
\begin{algorithmic}[1]
\STATE \textbf{Input:} Raw Inhibition Game Eye Tracking Data (CSV Files)
\STATE Parameters: Screen $(1920 \times 1080)$, tolerance = 50px
\STATE Parse coordinates: $(x_i, y_i) \leftarrow$ EyeTracker strings
\STATE Remove invalid data: missing coordinates, $(0,0)$ points, "Picked Trash" events
\STATE Validate bounds
\STATE Define AOIs
\STATE Create mapping: $f_{AOI}: (x,y) \rightarrow AOI_{name} \cup \{\emptyset\}$
\STATE \textbf{Output:} Clean Final dataset
\end{algorithmic}
\end{algorithm}

Algorithm~\ref{alg_eye_movements} implements a dual-threshold approach combining fundamental I-VT \cite{salvucci2000ivt,olsen2012tobii} classification with spatial clustering. This unified method provides both rapid classification and detailed fixation characterization, including duration, dispersion, and spatial properties.

\begin{algorithm}[!ht]
\caption{Unified Eye Movement Detection}
\label{alg_eye_movements}
\begin{algorithmic}[1]
\STATE \textbf{Input:} Clean data $D$, thresholds $V_{basic} = 721$px/s, $V_{advanced} = 300$px/s
\STATE \textbf{Parameters:} Spatial threshold $\tau = 50$px, min duration = 100ms
\STATE Initialize: $fixations \leftarrow \emptyset$, $saccades \leftarrow \emptyset$, $cluster \leftarrow \emptyset$

\FOR{$i = 2$ to $|D|$}
 \STATE $velocity \leftarrow \frac{\sqrt{(x_i - x_{i-1})^2 + (y_i - y_{i-1})^2}}{(t_i - t_{i-1})/1000}$
 
 \STATE \textbf{Basic Classification:} $type_i \leftarrow$ fixation if $velocity \leq V_{basic}$, else saccade
 
 \STATE \textbf{Advanced Clustering:}
 \IF{$velocity < V_{advanced}$ AND spatial distance to cluster $\leq \tau$}
 \STATE $cluster \leftarrow cluster \cup \{i\}$
 \ELSE
 \IF{$|cluster| \geq 3$ AND duration $\geq 100$ms}
 \STATE Calculate: center, duration, dispersion
 \STATE $fixations \leftarrow fixations \cup \{ProcessCluster(cluster)\}$
 \ENDIF
 \STATE Record saccade: amplitude, peak\_velocity, duration
 \STATE $cluster \leftarrow \{i\}$
 \ENDIF
\ENDFOR
\STATE \textbf{Output:} Fixations and saccades with properties

\end{algorithmic}
\end{algorithm}

\subsection{Performance Analysis and Behavioral Assessment}
Algorithm~\ref{alg_performance_behavioral} integrates task performance assessment through target-click matching and comprehensive attention analysis. Behavioral classification identifies visual processing strategies based on fixation duration and saccade amplitude patterns, providing insights into the efficiency of cognitive processing.

\begin{algorithm}[!ht]
\caption{Performance Analysis and Behavioral Assessment}
\label{alg_performance_behavioral}
\begin{algorithmic}[1]

\STATE \textbf{Input:} Eye movements, event timeline, AOI mapping

\STATE \textbf{Task Performance:}
\STATE Extract targets and clicks, initialize $matched \leftarrow 0$, $used \leftarrow \emptyset$
\FOR{each target $t$}
 \STATE Find unused click $c$ where $522 \leq (c.time - t.time) \leq 5000$ms
 \IF{valid click found}
 \STATE $matched \leftarrow matched + 1$, $used \leftarrow used \cup \{c\}$
 \ENDIF
\ENDFOR
\STATE $hit\_rate \leftarrow \frac{matched}{|targets|} \times 100$

\STATE \textbf{Attention Metrics:}
\STATE $attention\_scatter \leftarrow \sigma_x + \sigma_y$
\STATE $task\_relevance \leftarrow \frac{\text{gaze points in bins}}{\text{total gaze points}}$
\STATE $aoi\_transitions \leftarrow$ count transitions between AOIs
\STATE $gaze\_efficiency \leftarrow \frac{|fixations|}{|gaze\_points|}$

\STATE \textbf{Behavioral Classification:}
\STATE $avg\_fixation\_duration \leftarrow$ mean(fixation durations)
\IF{$avg\_fixation\_duration > 400$ms}
 \STATE $behavior \leftarrow$ "Deep processing/difficulty"
\ELSIF{$avg\_fixation\_duration < 200$ms}
 \STATE $behavior \leftarrow$ "Quick scanning"
\ENDIF

\STATE $avg\_saccade\_amplitude \leftarrow$ mean(saccade amplitudes)
\IF{$avg\_saccade\_amplitude > 300$px}
 \STATE $pattern \leftarrow$ "Broad visual search"
\ELSIF{$avg\_saccade\_amplitude < 100$px}
 \STATE $pattern \leftarrow$ "Focused examination"
\ENDIF

\STATE \textbf{Output:} Performance metrics and behavioral classification

\end{algorithmic}
\end{algorithm}

Algorithm~\ref{alg:risk_intervention} provides evidence-based risk stratification using validated thresholds for task focus (30\% critical), attention control (400 px scatter), and performance indicators. The system generates personalized interventions tailored to specific attention deficits and offers differentiated feedback for educational stakeholders.

\begin{algorithm}[!ht]
\caption{Risk Assessment and Personalized Intervention}
\label{alg:risk_intervention}
\footnotesize
\begin{algorithmic}[1]
\STATE \textbf{Input:} Performance metrics, attention indicators across levels
\STATE \textbf{Risk Assessment:}
\STATE Initialize: $risk\_score \leftarrow 0$, $factors \leftarrow \emptyset$
\IF{$task\_relevance < 0.30$}
 \STATE $risk\_score \leftarrow risk\_score + 3$, add "Critical task focus deficit"
\ELSIF{$task\_relevance < 0.50$}
 \STATE $risk\_score \leftarrow risk\_score + 2$, add "Low task focus"
\ENDIF

\IF{$attention\_scatter > 400$}
 \STATE $risk\_score \leftarrow risk\_score + 3$, add "Poor attention control"
\ENDIF

\IF{$aoi\_transitions > 60$}
 \STATE $risk\_score \leftarrow risk\_score + 2$, add "Hyperactive scanning"
\ENDIF

\IF{$hit\_rate < 50$}
 \STATE $risk\_score \leftarrow risk\_score + 3$, add "Very low performance"
\ENDIF

\STATE \textbf{Intervention Planning:}
\STATE $avg\_relevance \leftarrow$ mean across levels
\STATE Initialize: $interventions \leftarrow \emptyset$

\IF{$avg\_relevance < 40\%$}
 \STATE Add: "Focus training", "Reduce distractions", "Attention cuing"
\ELSIF{$avg\_relevance < 60\%$}
 \STATE Add: "Sustained attention practice", "Visual attention training"
\ENDIF
\IF{$risk\_score > 6$}
 \STATE $urgency \leftarrow$ HIGH, add "Immediate intervention needed"
\ELSIF{$risk\_score > 3$}
 \STATE $urgency \leftarrow$ MODERATE, add "Preventive measures"
\ENDIF
\STATE \textbf{Personalized Feedback:}
\FOR{each level}
 \IF{$task\_relevance \geq 70\%$}
 \STATE $performance \leftarrow$ "Excellent"
 \ELSIF{$task\_relevance \geq 50\%$}
 \STATE $performance \leftarrow$ "Good"
 \ELSE
 \STATE $performance \leftarrow$ "Needs Support"
 \ENDIF
\ENDFOR
\STATE Generate differentiated recommendations for students, teachers, and specialists
\STATE \textbf{Output:} Risk classification and personalized intervention plan
\end{algorithmic}
\end{algorithm}

\section{Experimental Analysis and Results}\label{sec_Experimental}
\begin{figure*}[!ht]
 \centering
 \subfloat[Level 1 Analysis]{
 \includegraphics[trim={0 0 0 1cm},clip,width=\textwidth]{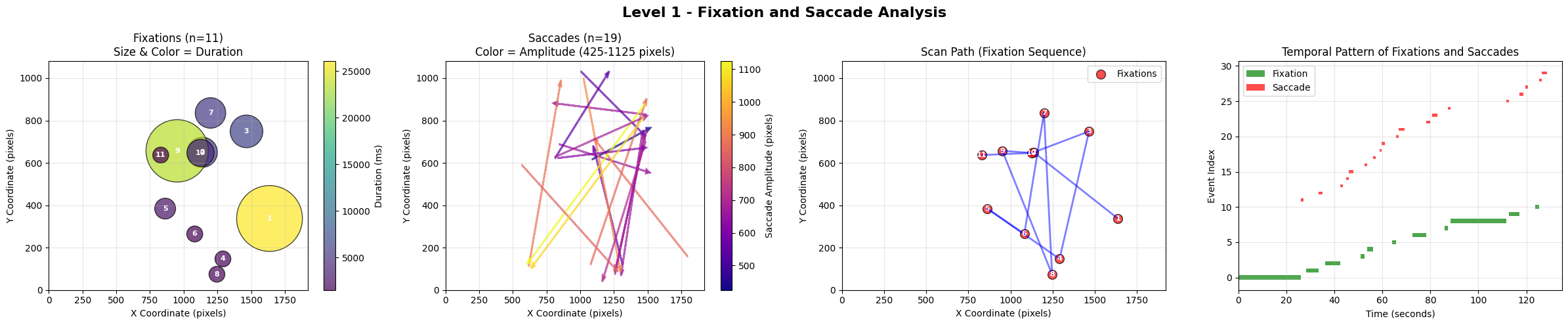}
 \label{fig:level1}
 }\\[1ex]
 \subfloat[Level 2 Analysis]{
 \includegraphics[trim={0 0 0 1cm},clip,width=\textwidth]{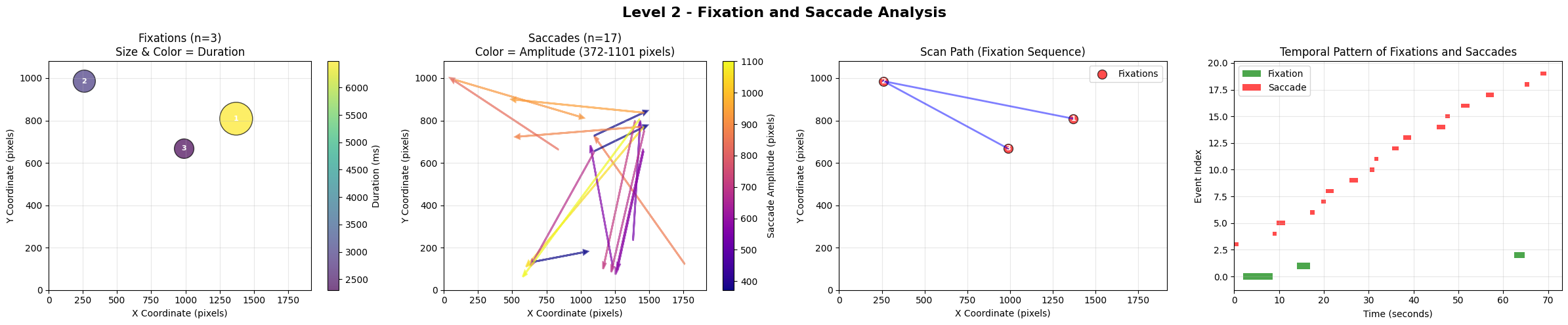}
 \label{fig:level2}
 }\\[1ex]
 \subfloat[Level 3 Analysis]{
 \includegraphics[trim={0 0 0 1cm},clip,width=\textwidth]{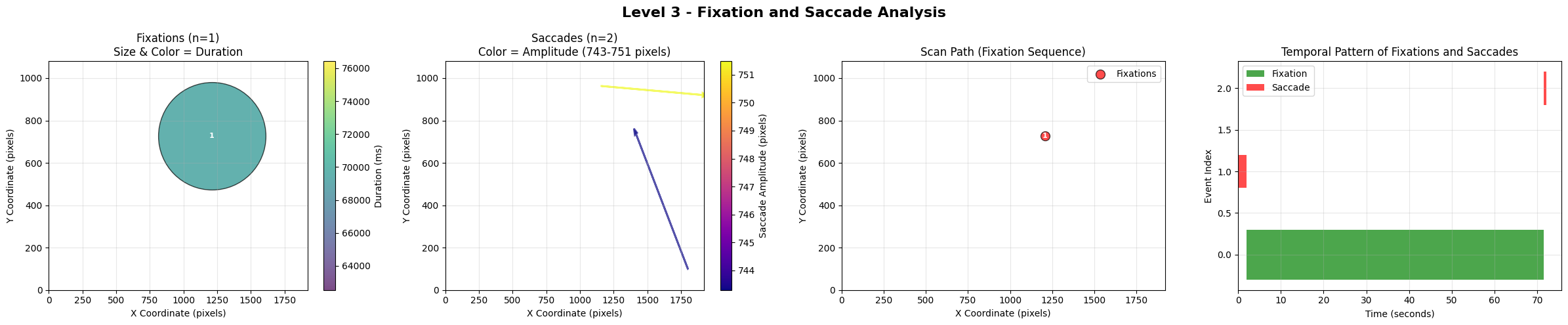}
 \label{fig:level3}
 }
 \caption{Fixation and Saccade analysis across three game levels: (a) Level 1, (b) Level 2, and (c) Level 3.}
 \label{fig:temporal_analysisdsdasdasd}
\end{figure*}

This section demonstrates the results of the proposed tool. 
%FIXATION ANALYSIS
The first plot, \ref{fig:level1}, \ref{fig:level2}, \ref{fig:level3}, from left to right, demonstrates where students fixated as colored circles. Larger, yellower circles indicate longer looking times at those locations. The fixation patterns reveal a notable progression as task difficulty increases from Level 1 to Level 3. Across the three levels, fixation behavior shows clear changes in cognitive engagement and visual attention distribution. At Level 1, the user made 11 fixations with a notably high average fixation duration of 7276.6 milliseconds, suggesting either deep cognitive processing or visual difficulty. Fixations were distributed across all four screen quadrants, but the \textbf{top right} region was the dominant focus. Level 2 showed a significant decrease in the number of fixations (3 total), with a lower average fixation duration of 3917.0 milliseconds, indicating a quicker visual assessment but still some cognitive effort, especially in the bottom right region, where most fixations occurred. In Level 3, fixation behavior became extremely concentrated: only one fixation was recorded, lasting an unusually long 69482.0 milliseconds, entirely in the \textbf{bottom right} quadrant. This suggests highly focused attention or even a possible moment of user inaction (e.g., reading or being cognitively overloaded).

%SACCADE ANALYSIS
The second plot, \ref{fig:level1}, \ref{fig:level2}, \ref{fig:level3}, from left to right, demonstrates Saccadic activity, showing how the user’s eyes moved between points of interest. In Level 1, there were 19 saccades with an average amplitude of 735.2 pixels and a peak velocity of 577.9 pixels/sec, indicating frequent and wide-ranging eye movements typical of thorough visual searching. Level 2 also had a high number of saccades (17), with a slightly lower average amplitude (707.1 pixels) and peak velocity (550.2 pixels/sec), still suggesting active scanning. Interestingly, in Level 3, saccades dropped to just 2, yet the average amplitude (747.4 pixels) and peak velocity (912.7 pixels/sec) remained high. This pattern suggests that, although eye movements were fewer, they were rapid and covered significant visual space, possibly moving directly to a point of focus before sustained attention took over.

%scan Path Analysis Across Difficulty Levels
The third plot \ref{fig:level1}, \ref{fig:level2}, \ref{fig:level3}, from left to right, maps out the sequence of eye movements, showing how gaze traveled from point to point across the game interface. It displays the sequential pattern of visual fixations (red circles numbered 1-11) connected by saccadic eye movements (blue lines) during an eye-tracking session. The graph reveals a non-systematic visual search strategy with fixations distributed across the interface, showing notable clustering in the middle-right region (coordinates around 1000-1300, 600-800), which suggests concentrated attention on task-relevant elements in that area. The irregular movement patterns, varying saccade lengths, and return visits to previously fixated locations indicate either exploratory behavior in an unfamiliar interface or difficulty locating specific targets, providing valuable insights into individual visual attention strategies and cognitive processing patterns that can inform educational interventions and interface design. Similarly, at level 2, this exploration requires 3 fixations and 2 saccades at a higher difficulty level, which may indicate highly efficient and targeted visual processing. The reduced number of fixations suggests the individual has developed a strategic, focused attention pattern where they can quickly identify and process task-relevant information without extensive visual exploration. However, at a higher difficulty level, this could also represent cognitive overload where the person is unable to engage in comprehensive visual search due to limited attentional resources, potentially leading to missed information or reduced task performance. With only one fixation at a higher difficulty level 3, this indicates severe attentional restriction and likely cognitive overload. The individual cannot effectively explore the visual scene, suggesting the task complexity has exceeded their processing capacity, leading to attentional tunneling or task disengagement. This pattern represents a critical warning sign requiring immediate educational intervention, as it indicates the difficulty level has overwhelmed the person's cognitive capabilities and prevented effective visual information processing.

%temporal pattern of fixation and saccades
The fourth plot in \ref{fig:level1}, \ref{fig:level2}, \ref{fig:level3}, from left to right, shows the temporal pattern, revealing that fixations (green bars) occurred more frequently in the later part of the viewing period, while saccades (red dots) were more spread out but also increased over time. Together, these visualizations reveal how visual attention was distributed across space and time during the experiment, with students spending more time focusing on specific areas and making more eye movements as the viewing session progressed. It can be noticed that at level 1, there were shorter saccadic movements and larger fixation times. At level 2, the fixation duration decreased rapidly. Similarly, at level 3, the duration decreased to a minimum, with one large bar representing the most extended duration.

The combined metrics further illustrate the shift in visual strategy across levels. Level 1 had a fixation/saccade ratio of 0.58, indicating a relatively balanced pattern of looking and moving, with a long total scan path of 13969.6 pixels, reflective of broad exploration. In Level 2, the ratio dropped sharply to 0.18, showing a high frequency of saccades relative to fixations, which aligns with a more exploratory or scanning approach; the scan path remained extensive at 12020.8 pixels. However, in Level 3, despite the high fixation duration, the fixation/saccade ratio rose to 0.50, and the total scan path shrank significantly to just 1494.8 pixels. This suggests a shift from broad scanning to focused attention, with minimal movement after an initial rapid eye movement. In all levels, long fixation durations point to cognitive effort, possibly due to task complexity or user uncertainty. The dominance of the *bottom right* region in Levels 2 and 3 suggests a spatial preference or that task-critical elements were located there. High saccade amplitudes across levels indicate the user was engaging in wide-ranging visual searches, although the decreasing number of fixations and scan path lengths suggests a narrowing focus. Particularly in Level 3, the combination of a prolonged single fixation and minimal movement implies the user concentrated intensely on one element, perhaps due to goal-directed behavior or a need for in-depth processing.

Table \ref{tab:cross_level} reveals the cross-level comparison showing changes in eye movement patterns as task difficulty increases, indicating progressive cognitive processing challenges. Level 1 demonstrates active visual exploration with 11 fixations and 19 saccades, suggesting systematic scanning behavior with moderate fixation durations (7276.6ms) and typical saccade characteristics. Level 2 shows a concerning reduction to only 3 fixations despite maintaining 17 saccades, resulting in the lowest fixation-to-saccade ratio (0.2), indicating fragmented attention with shortened processing periods (3917.0ms) but continued visual search activity. Level 3 exhibits extreme processing restriction with only 1 fixation and 2 saccades, but notably extended fixation duration (69482.0ms), suggesting either sustained focused attention on a single location or cognitive freezing where the individual becomes stuck due to task complexity overwhelming their processing capacity.
\begin{table*}[!ht]
\centering
\caption{Cross-Level Eye Movement Comparison}
\label{tab:cross_level}
\begin{tabular}{|c|c|c|p{2cm}|p{2cm}|p{2cm}|p{2cm}|}
\hline
\textbf{Level} & \textbf{Fixations} & \textbf{Saccades} & \textbf{Avg Fixation Duration (ms)} & \textbf{Avg Saccade Amplitude (px)} & \textbf{Avg Saccade Velocity (px/s)} & \textbf{Fix/Sacc Ratio} \\
\hline
1 & 11 & 19 & 7276.6 & 735.2 & 577.9 & 0.6 \\
\hline
2 & 3 & 17 & 3917.0 & 707.1 & 550.2 & 0.2 \\
\hline
3 & 1 & 2 & 69482.0 & 747.4 & 912.7 & 0.5 \\
\hline
\end{tabular}
\end{table*}

\begin{figure*}[!ht]
 \centering
 \subfloat[Level 1]{
 \includegraphics[trim={0 0 0 1cm},clip,width=\textwidth]{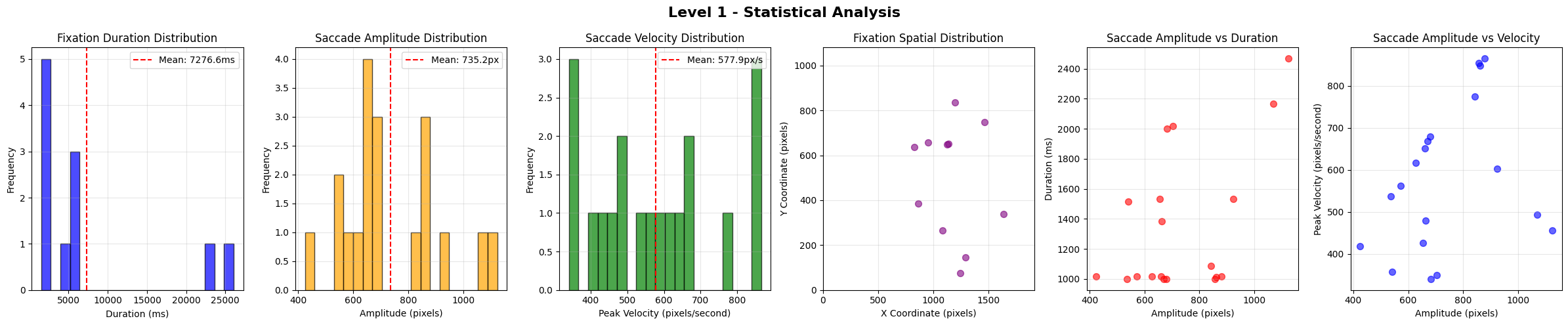}
 \label{fig:level1stat}
 }\\
 \subfloat[Level 2]{
 \includegraphics[trim={0 0 0 1cm},clip,width=\textwidth]{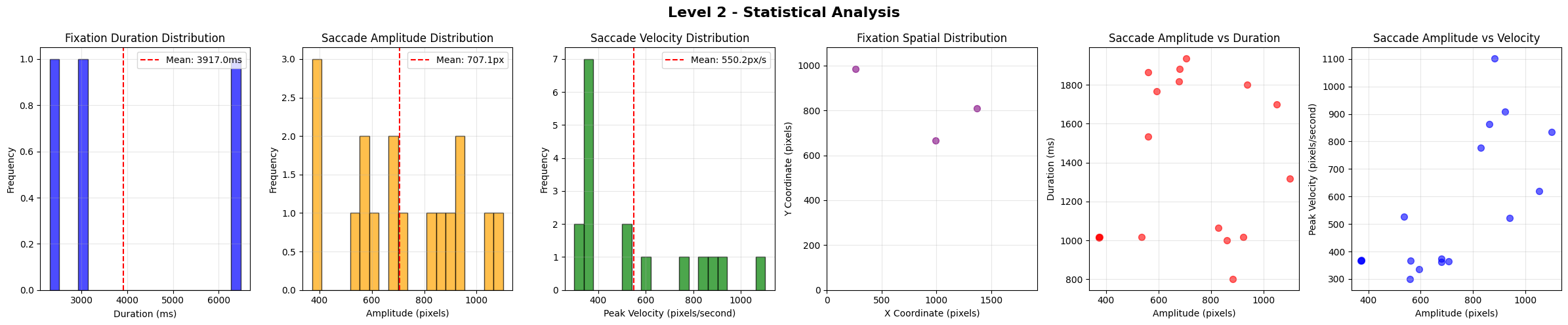}
 \label{fig:level2stat}
 }\\
 \subfloat[Level 3]{
 \includegraphics[trim={0 0 0 1cm},clip,width=\textwidth]{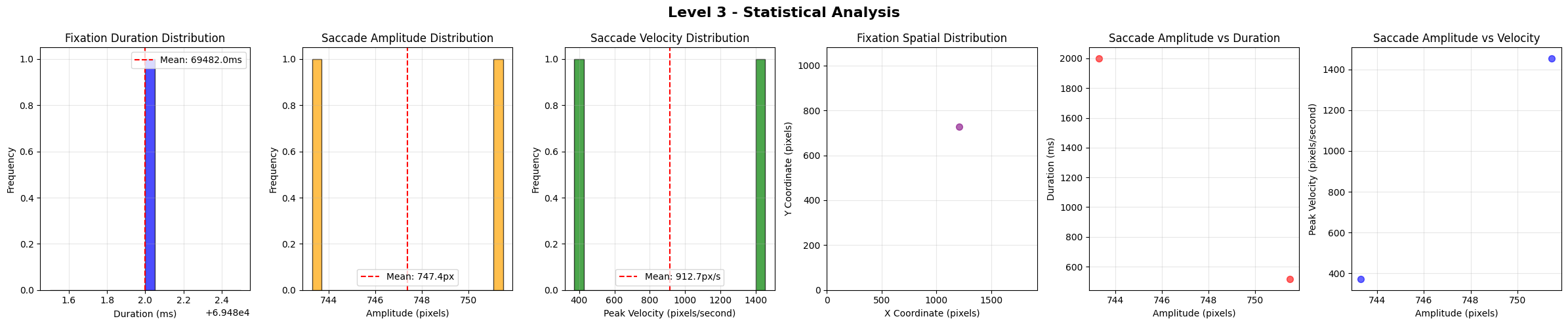}
 \label{fig:level3stat}
 }
 \caption{Statistical analysis across three game levels: (a) Level 1, (b) Level 2, and (c) Level 3.}
 \label{fig:temporal_analysis}
\end{figure*}

The fixation duration distribution in Fig. \ref{fig:level1stat} shows a right-skewed pattern (mean ~7276.6ms) with most fixations being brief but some extending longer, indicating varied processing demands. Saccade amplitude distribution (mean ~735.2px) displays broad variability, suggesting extensive visual exploration. Saccade velocity distribution centers around 577.3px/s with moderate spread. The fixation spatial distribution shows scattered points across the screen, indicating exploratory scanning. The amplitude vs duration scatter plot reveals diverse movement patterns.
In contrast, the amplitude vs velocity plot shows typical eye movement relationships with higher amplitudes generally corresponding to higher velocities, characteristic of initial task familiarization. In Fig. \ref{fig:level2stat}, Fixation durations at level 2 become more polarized (mean ~3917.0ms) with a bimodal distribution showing both quick recognition and deeper processing fixations. Saccade amplitudes decrease slightly (mean ~707.1px) with tighter clustering, indicating more targeted movements. Velocity distribution (mean ~550.2px/s) shows similar central tendency but reduced variability. Spatial fixations concentrate more on specific screen regions rather than scattered exploration. The amplitude-duration relationship becomes more structured, while the amplitude-velocity correlation shows emerging systematic patterns, suggesting developing visual efficiency and strategic scanning as users adapt to the task. In Fig. \ref{fig:level1stat}, the Fixation duration shows extreme concentration at level 3 with a single prominent peak (mean ~64827.0ms), indicating sustained focused attention on minimal locations. Saccade amplitude distribution becomes highly constrained (mean ~747.4px) with minimal spread. Velocity clustering tightens around 912.7px/s, showing very controlled movements. Spatial distribution demonstrates extreme concentration with minimal scatter across the screen. Both scatter plots show highly constrained relationships with few data points, reflecting either mastery-level efficiency with minimal visual exploration needed, or cognitive overload where increased difficulty severely restricts normal scanning behavior and limits comprehensive visual processing.

\begin{figure*}[!ht]
 \centering
 \includegraphics[trim={0 0 0 1cm},clip,width=\textwidth]{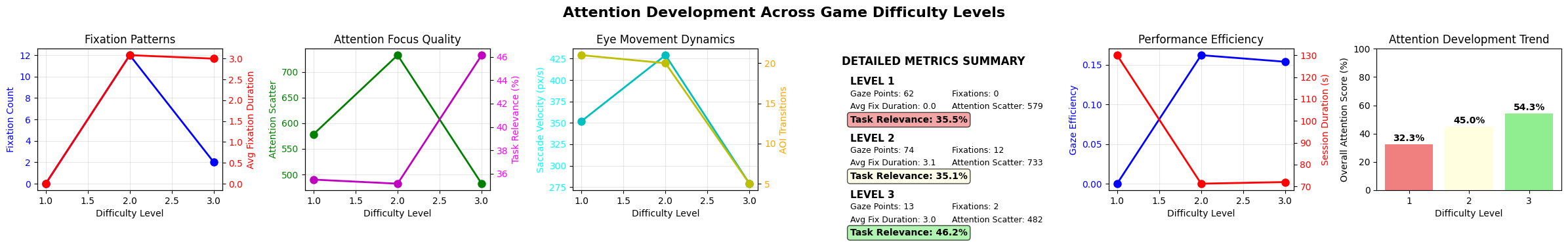}
 \caption{Attention Development Across All Levels}
 \label{Attentiondevelopment}
\end{figure*}

Fig. \ref{Attentiondevelopment} demonstrates attention development patterns across three difficulty levels in an educational eye-tracking assessment. The analysis reveals a concerning decline in attention quality as task complexity increases, with fixation counts rising from 0 to 12 fixations between levels 1 and 2 before dropping to 2 at level 3, while average fixation durations show extreme variability. Attention focus quality deteriorates significantly, with task relevance declining from initially stable performance to 35.1\% at level 2 and recovering slightly to 46.2\% at level 3, accompanied by increasing attention scatter (579 to 733 to 482 pixels). Eye movement dynamics show initially high saccade velocities and AOI transitions that peak at level 2 before declining at level 3, while performance efficiency drops precipitously from 0.15 to near-zero, indicating severe processing difficulties. The overall attention development trend scores reflect this decline (54.3\% to 45.0\% to 32.3\%), suggesting that increasing task difficulty overwhelms the individual's cognitive processing capacity, leading to fragmented attention, reduced task focus, and compromised visual processing efficiency rather than adaptive learning.

\begin{figure*}[!ht]
 \centering
 \includegraphics[trim={0 0 0 1cm},clip,width=\textwidth]{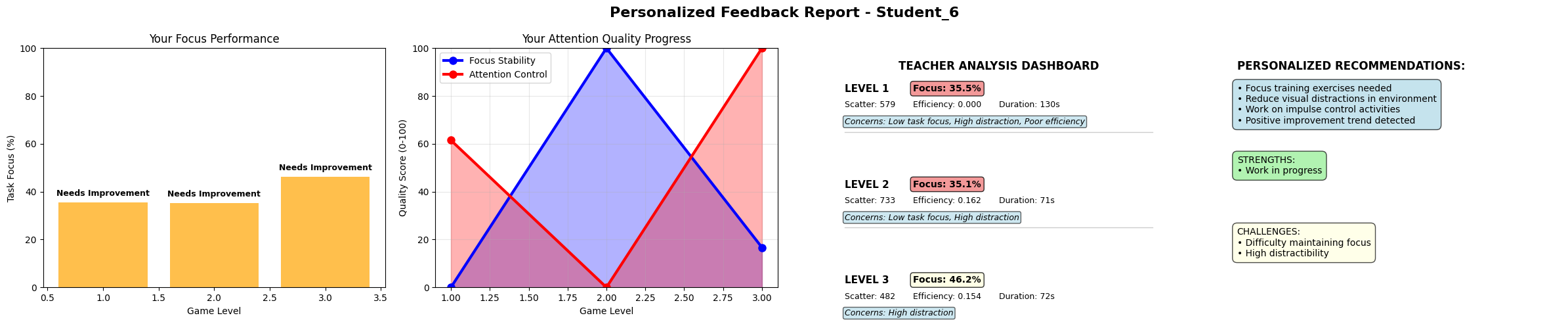}
 \caption{Personalized Feedback for Student 6}
 \label{Personalizedfeedback}
\end{figure*}

Fig. \ref{Personalizedfeedback} demonstrates a personalized feedback report for the student, revealing consistent attention difficulties across all three difficulty levels. Task focus performance remains in the "Needs Improvement" category (35.5\% to 35.1\% to 46.2\%) throughout the assessment. The attention quality progress chart reveals concerning patterns, where focus stability (blue line) initially starts low, peaks dramatically at level 2, and then drops significantly at level 3. In contrast, attention control (red line) follows an inverse pattern, indicating inconsistent and problematic attention regulation. The teacher analysis dashboard identifies persistent concerns, including low task focus, high distraction, and poor efficiency across all levels, with attention scatter remaining elevated (579 to 733 to 482 pixels) and processing efficiency near zero. The system generates targeted recommendations, including focus training exercises, environmental distraction reduction, and impulse control training. It identifies "difficulty maintaining focus" and "high distractibility" as primary challenges, with "work in progress" noted as a strength. This suggests that the student requires immediate, comprehensive attention, intervention and support.

\begin{figure*}[!ht]
 \centering
 \includegraphics[trim={0 0 0 1cm},clip,width=\textwidth]{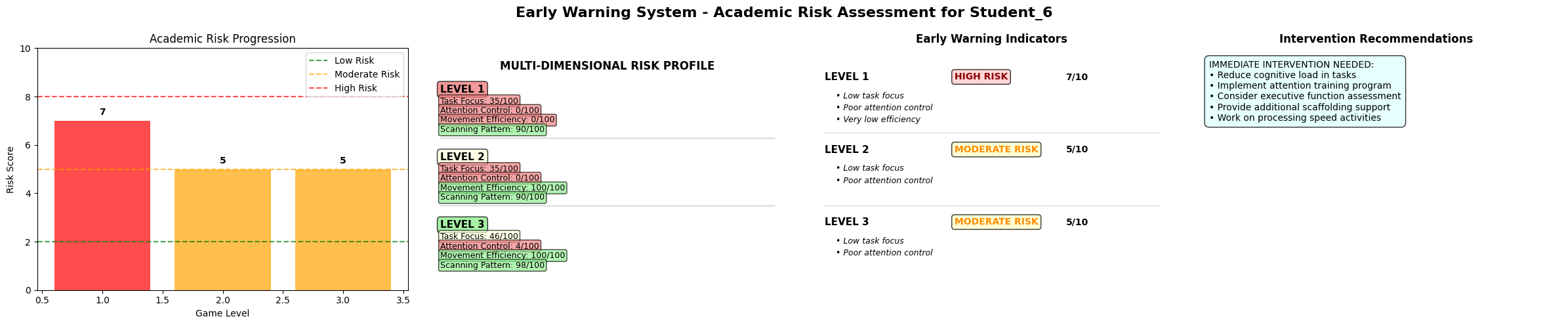}
 \caption{Early Warning and Risk Assessment}
 \label{figEarlyWarning}
\end{figure*}

Fig. \ref{figEarlyWarning} demonstrates that early warning system assessment reveals critical academic risk patterns requiring immediate intervention, with an initial HIGH RISK classification (7/10) at Level 1 that moderates to 5/10 at Levels 2 and 3. The multi-dimensional risk profile shows consistently poor performance across all assessed domains, with task focus scores remaining critically low (35-46/100), attention control severely impaired (0-41/100), movement efficiency near zero (0-10/100), and scanning patterns highly problematic (90-98/100), indicating fundamental attention regulation difficulties. The early warning indicators identify persistent risk factors, including low task focus, poor attention control, and very low efficiency across all levels, with Level 1 showing the most severe impairment. The system generates immediate intervention recommendations, including cognitive load reduction, attention training program implementation, executive function assessment, and processing speed activities, reflecting the urgent need for comprehensive educational support to address this student's significant attention and cognitive processing deficits before they further impact academic progress.

\section{Conclusion and Future Work}\label{sec_conclusion}
This eye-tracking analysis system successfully demonstrates the practical application of objective gaze measurement for educational assessment and intervention planning. The comprehensive algorithmic framework effectively processes raw eye-tracking data into meaningful educational insights through unified fixation detection, performance analysis, and personalized feedback generation. The case study with Student\_6 revealed critical attention regulation difficulties requiring immediate intervention, with consistently low task focus, elevated attention scatter, and processing efficiency deficits across difficulty levels. The system's multi-dimensional risk assessment accurately identified HIGH to MODERATE risk classifications, generating targeted interventions including focus training, attention regulation support, and instructional modifications. The early warning capabilities enable proactive educational support before attention deficits significantly impact academic progress. The evidence-based approach provides objective metrics that complement traditional assessment methods, supporting personalized learning strategies and data-driven educational decision making. This technology represents a valuable advancement in educational assessment, offering educators and specialists concrete tools for identifying at-risk students and implementing evidence-based interventions to support optimal learning outcomes.

Future work should focus on developing adaptive difficulty adjustment algorithms that maintain optimal cognitive load during assessment, and expanding the system to include real-time intervention delivery that provides immediate feedback and support based on detected attention patterns. Additionally, longitudinal studies across diverse populations and learning contexts are needed to establish normative benchmarks and validate the system's effectiveness in improving educational outcomes through personalized attention training interventions.

\bibliographystyle{ACM-Reference-Format}
\bibliography{ref-inhib}
\end{document}